# Protein Folding as a Quantum Transition Between Conformational States

Liaofu Luo*

Laboratory of Theoretical Biophysics, Faculty of Physical Science and Technology,

Inner Mongolia University, Hohhot 010021, China

*Email address: lolfcm@mail.imu.edu.cn

**Abstract**
The importance of torsion vibration in the transmission of life information is indicated. The localization of quantum torsion state is proved. Following these analyses a formalism on the quantum theory of conformation-electron system is proposed. The conformational-electronic transition is calculated by non-adiabatic operator method. The protein folding is viewed from conformational quantum transition and the folding rate is calculated. The time-scale of microsecond to millisecond for the fundamental folding event (nucleation, collapse, etc) is deduced. The dependence of transition rate W on N inertial moments is given. It indicates how W increases with the number N of torsion angles and decreases with the inertial moment I of atomic group in cooperative transition. The temperature dependence is also deduced which is different from chemical reaction in high-temperature region. It is demonstrated that the conformational dynamics gives deep insights into the folding mechanism and provides a useful tool for analyzing and explaining experimental facts on the rate of protein folding.

## 1. Introduction

One of the most important challenges in biology is to predict the three-dimensional structure of protein from its primary amino acid sequence. Protein folding is a complex process which may be influenced by many factors, for example, the occurrence of transition state ensemble, the slow reshuffling reactions of almost native-like intermediates, the association of prefolded monomers and the formation of disulfide bonds, the existence of enzymes as catalysts and molecular chaperones which prevent improper protein aggregation *in vivo*, etc. The complexity inherent in the many-body system, which is composed of thousands of atoms and millions of possible inter-atomic interactions, makes the structure prediction nearly impossible. However, the fundamental physics underlying folding may be much simpler than this complexity [1]. Protein folding rate is one of the basic aspects closely related to folding mechanism and one of the important experimental measurements that have been made about folding process [2]. The speed of protein folding takes various values which changes mainly from miliseconds to an hour. Some proteins fold within microseconds [3]. Recently, great progress has been made on the prediction of protein folding speed from the effective contact order of residues and other related measures[4-8].

The nucleation condensation mechanism on the rates of protein folding was proposed [9]. The nucleus is composed of a set of adjacent residues and stabilized by long-range interactions. The latter is formed as the rest of the protein collapses around the nucleus. In nucleation condensation mechanism the "first principle prediction" of protein folding rates was given. The approach consists of two steps: the random, diffusive sampling to find the native tertiary topology and the local conformational changes within the native topology to find the unique native state. The first step is assumed to be the rate-limiting one which obeys Einstein diffusion equation and the intra-protein diffusion constant was determined experimentally [10]. The diffusion control in an elementary protein folding reaction was also discussed in literatures [11]. However, in all these studies some fundamental problems on protein folding mechanism are still open and need to be clarified. For example, how to compute nucleation rate of the nucleus formation? How to estimate the time needed for the collapses around the nucleus? How to calculate the time of a pair of residues to form a contact with respect to residue distance? How to make a theoretical calculation on diffusion constant from the interaction of conformational changes in adjacent residues? Evidently, to solve these problems a more fundamental standpoint should be taken. Following our point the folding is a self-organization process which consists of many steps of conformational transitions and the quantum transitions between a group of conformational states are the elementary processes of protein folding. As a primary study we shall propose a quantum transition theory of protein folding in this article and calculate the rate of the basic step of the folding. We shall discuss what is the time scale of the elementary folding event in the complex folding process.

Two problems on conformation dynamics of biological macromolecules will be discussed at first [12], one is related to the torsion vibration and the definition of conformational state, and the other is conformational quantum transition (section 2 and 3). Then we shall discuss the time scale of protein folding based on the conformation dynamics of macromolecules. The rate of conformational transition will be deduced from the nonadiabaticity operator method. We shall show that the result can be used to explain the time scale of microsecond to milisecond for protein folding as observed in experiments. Some interesting relations of the transition rate dependent on various physical quantities will also be deduced (section 4 and 5).

## 2. Torsion vibration and the localization of conformational state

There are huge numbers of variables in a biological system. What are the fundamental variables in the life processes at the molecular level? Since the classical works of B. Pullman and A. Pullman (1963) on nucleic acids, it is generally accepted that the mobile $\pi$ electrons play an important role in the biological activities of macromolecules. On the other hand, the increasing importance of conformational variations for biological functions of biopolymers produced by rotations of atomic groups around a single bond has been realized. Therefore, it is reasonable to assume that the main variables in life processes at the molecular level are the conformation of biological macromolecules and their frontier electrons.

A macromolecule consisting of n atoms (if an atom is looked as one point）includes 3n coordinates. Apart from 6 translational and rotational degrees of freedom there are 3n-6 coordinates describing molecular shape. It has been proved that the bond lengths, bond angles and torsion (dihedral) angles form a complete set to describe the molecular shape. The energy related to the variation of bond length and bond angle is in the range of 1-7 Kcal/mole, while that related to torsion vibration is 0.1-1 Kcal/mole, much lower than that of stretching and bending modes. Comparing with the average thermal energy in room temperature ($25^0C$) 0.59 Kcal/mole or 0.026 ev, we know that the torsion angles are easily changed and they, together with frontier electron coordinates, consist of the main variables for describing a living system.

The torsion energy 0.1-1 Kcal/mole corresponds to vibration frequency $10^{12}-10^{13}$ Hz in the range of microwave. From the vibrational partition function of a molecule in harmonic conformational potential.

$$Z = (e^{(1/2)\beta\hbar\omega} - e^{-(1/2)\beta\hbar\omega})^{-1} \tag{1}$$

one deduces the average energy $\bar{E}$ and entropy $S$ readily:

$$\bar{E} = \frac{\hbar\omega}{2} + \hbar\omega(e^{\beta\hbar\omega} - 1)^{-1} \tag{2}$$

$$S = k_B\{\beta\hbar\omega(e^{\beta\hbar\omega} - 1)^{-1} - \ln(e^{\beta\hbar\omega} - 1) + \beta\hbar\omega\} \tag{3}$$

For a molecule with many torsion angles the potential can be expanded into harmonic modes and each mode contributes to internal energy and entropy of the molecule described by (2) and (3). As is well known, the entropy is related to information quantity by

$$S = (k_B \ln 2)I \tag{4}$$

Thus, by means of (3) one sees the strong dependence of information quantity on frequency (Table 1, $T = 300$ K). As the frequency $\omega$ is lower than $10^{13}$ Hz, the information quantity $I$ increases rapidly with the decreasing $\omega$. This is the so-called Boson condensation. Sine the typical value of torsion vibration is $10^{12}$–$10^{13}$ Hz, the torsion vibration may play an important role in the transmission of information for biological macromolecules.

Table1. The dependence of information quantity $I$ on frequency $\upsilon$ (T=300K).

| $v$ | $10^{14}$ | $10^{13}$ | $10^{12}$ | $10^{11}$ | $10^{10}$ | $10^{9}$ | $10^{8}$ |
|---|---|---|---|---|---|---|---|
| $I$ | ~$10^{-6}$ | 0.63 | 2.83 | 4.14 | 6.44 | 8.74 | 11.04 |

To study conformational dynamics from the point of quantum mechanics we should discuss the localization of conformational state at first.

Suppose that the conformation is described by a set of torsion angles $\{\theta\} = (\theta_1 \ldots \theta_s)$. The conformation potential $U$, as a function of $\theta$, generally has some minima that correspond to stable conformations. To be definite, consider the case of two minima. Set

$$U(\theta) = \begin{cases} \frac{\kappa}{2}(\theta + \frac{\pi}{2})^2 & [-\pi, 0] \\ \frac{\kappa}{2}\{(1-\lambda)(\theta - \frac{\pi}{2})^2 + \frac{\pi^2}{4}\lambda\} & [0, \pi] \end{cases} \tag{5}$$

Here $\lambda$ is an asymmetrical parameter. We shall discuss the dependence of steady–state solutions on the asymmetrical parameters and demonstrate that the drawn conclusions will not depend on the particular choice of the form of potential. If $\lambda = 0$, then $U(\theta)$ is $C_2$ symmetrical, and the conformational wave functions should be nonlocal and no definite conformation can be related with the molecule. However if $\lambda \neq 0$, the $C_2$–symmetry is broken; by numerical solution of Schrodinger equation for the potential given by (5), one can show that for $I\kappa/\hbar^2 = 10$ ($I$ is the inertia moment of the molecule with respect to the coordinate $\theta$), the ground state would be localized to an amount larger than 90% as $\lambda \geq 10^{-3}$, and for $I\kappa/\hbar^2 = 40$, the ground state would be localized to larger than 99% as $\lambda \geq 10^{-6}$ [12]. Define the locality $J_i$ of the $i$-th level as

$$J_i = \frac{\max(P_i, Q_i)}{\min(P_i, Q_i)} \tag{6}$$

where $P_i$ is the probability of $i$-th level located in the "left" conformation ($-\pi \leq \theta \leq 0$) and $Q$ in the "right" conformation ($0 \leq \theta \leq \pi$). By numerical calculation one can show that nearly all bound states are localized with $J_i > 2$ as $\lambda \geq 0.1$ (0.01) for the case of $I\kappa/\hbar^2 = 10$ (40). When $\lambda$ small ($\lambda \leq 0.2$ for $I\kappa/\hbar^2 = 10$ and $\lambda \leq 0.1$ for $I\kappa/\hbar^2 = 40$) the left and right energy valleys are occupied alternately from ground to excited levels. However for large $\lambda$, the filling number in left valley is increased.

The above discussion shows that the small asymmetry in potential would cause the strong localization of wave functions. Although the results are deduced from a particular potential (5), the generality of the conclusions can be demonstrated by the following argument. Consider the interaction between a pair of levels localized in left potential valley $A$ (the corresponding wave function denoted as $\psi_A$) and right potential valley $B$ (the corresponding wave function $\psi_B$) respectively. Set Hamiltonian

$$H = \begin{bmatrix} h_a & g \\ g & h_b \end{bmatrix} \tag{7}$$

corresponding to

$$E = E_\pm = \frac{1}{2}\{(h_a + h_b) \pm [(h_a - h_b)^2 + 4g^2]^{1/2}\} \tag{8}$$

The localized wave functions

$$x_+ = \begin{pmatrix} 0 \\ 1 \end{pmatrix}, \quad x_- = \begin{pmatrix} 1 \\ 0 \end{pmatrix} \quad \text{(for } h_b > h_a\text{)}$$

or

$$x_+ = \begin{pmatrix} 1 \\ 0 \end{pmatrix}, \quad x_- = \begin{pmatrix} 0 \\ 1 \end{pmatrix} \quad \text{(for } h_b > h_a\text{)} \tag{9}$$

are obtained if

$$|g| \ll |h_a - h_b| \tag{10}$$

or its equivalent

$$\text{overlap integral } |\int \psi_A^* \psi_B d\theta| \ll \text{asymmetry } \lambda \tag{10a}$$

Therefore, given the asymmetry parameter $\lambda$, a pair of wave functions satisfying Eq. (10) are localized. Since the overlap integral of ground states is small, Eq. (10) is easily satisfied even for a very small asymmetry in potential, and the localization of the ground state occurs. To estimate the overlap integral, set

$$\frac{1}{\hbar}\{2I[U(\theta) - E]\}^{1/2} = K(\theta) \tag{11}$$

From the WKB approximation the wave functions in the barrier region are

$$\begin{aligned}\psi_A &= N_A K(\theta)^{-1/2} \exp\{-\int_\alpha^\theta K(\theta')d\theta'\} \\ \psi_B &= N_B K(\theta)^{-1/2} \exp\{-\int_\theta^\beta K(\theta')d\theta'\}\end{aligned} \qquad (\alpha < \theta < \beta) \tag{12}$$

$\alpha$ and $\beta$ are turning points at the left and right sides of the barrier, respectively ($U(\alpha) = U(\beta) = E$). Inserting (12) into (10a), one obtains

$$N_A N_B \int_\alpha^\beta d\theta K(\theta)^{-1} \exp\{-\int_\alpha^\beta K(\theta')d\theta'\} \ll \lambda \tag{13}$$

Therefore, the condition of localization is

$$\exp\{-\int_\alpha^\beta K(\theta')d\theta'\} \ll \lambda \tag{14}$$

For example, set $\beta - \alpha \sim 1$, $U - E \sim 0.5$ Kcal/mol; then $\lambda \gg 10^{-13}$ for $I = 10^{-38}$ g·cm$^2$ and $\lambda \gg \frac{1}{150}$ for $I = 3 \times 10^{-40}$ g·cm$^2$. In fact, the conformational potential usually has certain asymmetry, so the levels below the top of barrier are always localized apart from those near the top where $K(\theta) \sim 0$. As for the problem whether the ground state is localized in the left or right valley, it is determined by the comparison of the depths of two valleys and the values of $\partial^2 U / \partial \theta^2$ in two minima of the potential. As Anderson studied the excitations on a disordered lattice, he proposed a noted theorem: the eigenfunctions are localized if the strength of the disorder exceeds some definite value [13]. Now, the theory of Anderson localization has been generalized to the system of molecular conformations.

We have discussed the possibility of the definition of a localized conformational state. Now we will point out the peculiarities of the spectrum of conformational vibration. For simplicity, suppose a periodic potential with period $2\pi$ and it taking the form

$$U(\theta) = \frac{1}{2} I \omega^2 (\theta - \pi)^2 \tag{15}$$

between $0 \le \theta \le 2\pi$. By numerical solution of the Schrodinger equation, one proves

$$E = E_n = (n + \frac{1}{2})\hbar\omega \qquad (E < \frac{1}{2} I\pi^2 \omega^2) \tag{16}$$

When $E \sim (1/2) I \pi^2 \omega^2$, there is explicit deviation from (16). Suppose $x$ level near $(1/2) I \pi^2 \omega^2$ that deviate from (16) explicitly. For these levels, the wave functions in $(0, 2\pi)$ overlap with those in $(-2\pi, 0)$ at $\theta \sim 0$, and with those in $(2\pi, 4\pi)$ at $\theta \sim 2\pi$. At $\theta$ near 0 the overlap integral is determined by

$$\exp(-\int_{-\alpha/2}^{\alpha/2} \frac{1}{\hbar}\{2I[U(\theta) - \frac{1}{2}I\pi^2\omega^2 + x\hbar\omega]\}^{1/2} d\theta) \simeq \exp(-\frac{\alpha}{\hbar}\{2Ix\hbar\omega\}^{1/2}) \tag{17}$$

where $\alpha$ is related to $x$ through

$$\frac{1}{2}I\pi^2\omega^2 - x\hbar\omega = \frac{1}{2}I\omega^2(\pi - \frac{\alpha}{2})^2$$

or

$$\alpha = \frac{2x\hbar}{\pi I \omega} \tag{18}$$

Inserting (18) into (17), one sees that the condition of explicit overlap of wave functions is

$$\frac{2}{\hbar}(2Ix\hbar\omega)^{1/2} = (\frac{8\hbar}{\pi^2 I \omega}x^3)^{1/2} = 1 \tag{19}$$

On the other hand, below $(1/2)I\pi^2\omega^2$ there are $N$ levels

$$N = \frac{\pi^2 I \omega}{2\hbar} \tag{20}$$

From (19) and (20) we obtain

$$x^3 = \frac{1}{4}N \tag{21}$$

For example, taking $I = 10^{-38}$ g cm$^2$, $\omega = 6 \times 10^{12}$ Hz, then $N = 300$ and $x = 4$. That is, only four levels near the top of barrier deviate from the equal–spacing law. The numerical solution of the Schrodinger equation has verified the above result. Next, we will consider the spectrum higher than $(1/2)I\pi^2\omega^2$. If the level is high enough, then the potential will be neglected as compared with kinetic energy. The spectrum is of the free rotator type. It can be shown that there exist double degeneracies for $n = 2m$ and $2m+1$, namely

$$E = E_n = \frac{\hbar^2 n^2}{2I} \qquad (n = 2m, 2m+1) \qquad (E > \frac{1}{2}I\pi^2\omega^2) \tag{22}$$

The degeneracy is easily understood by introducing an operator of parity

$$\Pi \psi(\theta) = \psi(2\pi - \theta) \tag{23}$$

Because $\Pi$ is commutated with Hamiltonian, an energy level of parity even is degenerate with the corresponding one of parity odd. In conclusion, owing to the periodic boundary condition, the spectrum of harmonic conformational potential is of equal spacing only when $E < (1/2)I\pi^2\omega^2 - xh\omega$. Near the top of the barrier [$E \sim (1/2)I\pi^2\omega^2$] the spectrum is distorted, and above the top it gives a double-degenerate rotator-like spectrum.

## 3. Conformational-electronic transition calculated by non-adiabatic operator method

The conformation variation of a macromolecule is a process of quantum transition between vibrational states of different conformations. In general, the transition is closely related to the electronic states of the molecule. The quantum mechanics of a conformation–electron system is described by conformation electron field theory. For simplicity, consider the case of a single frontier electron. The dynamical variables of the system are ($\theta, x$) ($x$ is the coordinate of the frontier electron and θ the torsion angle of molecule), and the wave function $M(\theta, x)$ satisfies [12]

$$(H_1(\theta, \frac{\partial}{\partial \theta}) + H_2(\theta, x, \nabla))M(\theta, x) = EM(\theta, x) \tag{24}$$

Because the mass of electrons is much smaller than nucleic masses, the adiabatic approximation can be used. Though the velocity of nuclei may not be small in the course of conformational change, we need only discuss the initial and final states of the transition. Thus the approximation will not bring too many errors (only of the order of $m_e/m_n$). In adiabatic approximation the wave function can be expressed as

$$M(\theta, x) = \psi(\theta)\varphi(x, \theta) \tag{25}$$

and these two factors satisfy

$$H_2(\theta, x, \nabla)\varphi_\alpha(x, \theta) = \varepsilon^\alpha(\theta)\varphi_\alpha(x, \theta) \tag{26}$$

$$\{H_1(\theta, \frac{\partial}{\partial \theta}) + \varepsilon^\alpha(\theta)\}\psi_{kn\alpha}(\theta) = E_{kn\alpha}\psi_{kn\alpha}(\theta) \tag{27}$$

here $\alpha$ denotes the electron state, and $(k, n)$ refer to the conformational and vibrational state, respectively. Now we discuss transition operators in different cases.

We consider the case of absence of external field at first, which includes the protein folding, the radiationless electron transfer and the multiphonon relaxation in biological system, etc.. Because $M(\theta, x)$ is not a rigorous eigenstate of Hamiltonian $H_1 + H_2$, there exist transitions between adiabatic states that result from the off–diagonal elements

$$\int M^+_{k'n'\alpha'}(H_1 + H_2)M_{kn\alpha}d\theta d^3x = E_{kn\alpha}\delta_{kk'}\delta_{nn'}\delta_{\alpha\alpha'} + \langle k'n'\alpha' | H' | kn\alpha \rangle \tag{28}$$

$$\langle k'n'\alpha' | H' | kn\alpha \rangle = \int \psi^+_{k'n'\alpha'}(\theta)\{-\frac{\hbar^2}{2I}\int \varphi^+_\alpha(\frac{\partial^2 \varphi_\alpha}{\partial \theta^2} + 2\frac{\partial \varphi_\alpha}{\partial \theta}\frac{\partial}{\partial \theta})d^3x\}\psi_{kn\alpha}(\theta)d\theta \tag{29}$$

Here $H'$ is a Hamiltonian describing conformational transition. We see that the conformational transition is partly related to the electronic wave motion in the molecule and determined by the $\theta$ dependence of the electronic wave function $\varphi_a(x, \theta)$. Equation (29) is the generalization of the nonadiabaticity operator of Huang and Rhys in solid state physics (1950) [14]. In molecular orbital theory the electronic wave function can be expressed as the linear combination of atomic orbits, and the combination coefficients and $\varepsilon^a(\theta)$ can be obtained by solving Huckel equations.

Note: The quantum system with conformational variable as slow-varying variable is not only encountered in biological study, but also of special importance in general quantum theory. Berry solved the eigenvalue problem of this system rigorously and found the terms $\frac{\partial \varphi_\alpha}{\partial \theta}\frac{\partial}{\partial \theta}$ in the form of guage potential

$$A = \int \varphi^+_\alpha i \frac{\partial}{\partial \theta}\varphi_\alpha d^3x \tag{30}$$

occurring in the equation of conformational wave function $\psi(\theta)$. It leads to an additional phase factor $\exp\{-i\int Ad\theta\}$ in the solution of $\psi(\theta)$. This is so-called Berry phase which originates from the influence of conformational motion to electronic motion [15].

The nonadiabatic matrix element (29) will be simplified in two important cases.

*Case* 1.

For most protein folding problem the electronic state does not change in transition processes, namely $\alpha' = \alpha$. Because the wave function $\varphi_a$ is generally real, one can deduce

$$\int \varphi_a(x,\theta) \frac{\partial \varphi_a(x,\theta)}{\partial \theta} d^3x = 0 \qquad (31)$$

from the normalization condition $\int \varphi_a(x,\theta)\varphi_a(x,\theta)d^3x = 1$. Therefore, only the first term in Eq. (29) should be retained, namely

$$\langle k'n'\alpha | H' | kn\alpha \rangle = \int \psi^+_{k'n'\alpha}(\theta)\{-\frac{\hbar^2}{2I}\int \varphi^+_\alpha \frac{\partial^2 \varphi_\alpha}{\partial \theta^2} d^3x\}\psi_{kn\alpha}(\theta)d\theta \qquad (32)$$

The case will be discussed in detail in next section, section 4.

*Case* 2

When the perturbation approximation is valid, the $\theta$ dependence of electronic wave function $\phi_\alpha(x,\theta)$ can be deduced by the perturbation method.

$$H_2(\theta, x, \nabla) = H_2(\theta_0, x, \nabla) + (\frac{\partial H_2(\theta, x, \nabla)}{\partial \theta})_0 (\theta - \theta_0)$$
$$\equiv H_2(\theta_0, x, \nabla) + h(x, \nabla)(\theta - \theta_0) \qquad (33)$$

$$\varphi_\alpha(x,\theta) = \varphi_\alpha(x,\theta_0) + (\theta - \theta_0)\sum_\beta \frac{h_{\beta\alpha}}{\varepsilon^0_\alpha - \varepsilon^0_\beta}\varphi_\beta(x,\theta_0)$$

Inserting (33) into (29), only the second term is retained:

$$\langle k'n'\alpha' | H' | kn\alpha \rangle = -\frac{\hbar^2}{I}\int \varphi^+_{\alpha'}(x,\theta_0)\sum_\beta \frac{h_{\beta\alpha}}{\varepsilon^0_\alpha - \varepsilon^0_\beta}\varphi_\beta(x,\theta_0)d^3x \int \psi^+_{k'n'\alpha'}(\theta)\frac{\partial}{\partial \theta}\psi_{kn\alpha}(\theta)d\theta$$
$$(34)$$

The detailed calculation can be found in ref [12].

Next we will discuss the conformational change in the external field. There are two cases frequently occurred in biological systems, namely, the photon induced transition (Case 3) and the static electric field-induced transition (Case 4).

*Case* 3

For radiation transition the electromagnetic interaction can be introduced from the gauge invariance of $H_1 + H_2$ (Eq (24)). The deduced interaction is

$$H_{EM} = -\frac{ie\hbar}{mc}A\cdot\nabla + \frac{e^2}{2mc^2}A^2 \qquad (35)$$

$$A = \sum_{q,\lambda}(\frac{c^2\hbar}{2\Omega V_0})^{1/2}[e_\lambda c_{q\lambda}\exp i(q\cdot x - \Omega t) + e^*_\lambda c^+_{q\lambda}\exp -i(q\cdot x - \Omega t)]$$
$$(36)$$

For electron transition absorbing one photon, the matrix element

$$\langle k'n'\alpha' | H_{EM} | kn\alpha, q\lambda \rangle = \frac{F_E \exp i(E_{k'n'\alpha'} - E_{kn\alpha} - \hbar\Omega)t}{\hbar}$$

$$F_E = -\frac{ie}{m}(\frac{\hbar^3}{2\Omega V_0})^{1/2} \mathbf{e}_\lambda \cdot \int d^3 x \varphi_{\alpha'}^+(x, \theta_0) \nabla \varphi_\alpha(x, \theta_0)$$

$$\times \int d\theta \psi_{k'n'\alpha'}^+(\theta) \psi_{kn\alpha}(\theta)$$

$$-\frac{ie}{m}(\frac{\hbar^2}{2\Omega V_0})^{1/2} \mathbf{e}_\lambda \cdot (\frac{\partial}{\partial \theta}\int d^3 x \varphi_{\alpha'}^+(x, \theta) \nabla \varphi_\alpha(x, \theta))_{(\theta=\theta_0)}$$

$$\times \int d\theta \psi_{k'n'\alpha'}^+(\theta)(\theta - \theta_0) \psi_{kn\alpha}(\theta) \quad (37)$$

Here the perturbation expansion (33) can be used for the simplification of the second term of (37). The first term in the right–hand side (37) represents the Franck–Condon (FC) vertical transition, whereas the second term involving $\int d\theta \psi^+(\theta)(\theta - \theta_0)\psi(\theta)$ is the correction to the FC transition.

*Case* 4

The next case we will discuss is the static electric field–induced transition. Owing to the couplings of charged residues of a macromolecule with a Coulomb field, there exists an interaction $H_E(\theta)$ that results from the sum of Coulomb interactions of these residues with the external field. Without loss of generality, set $\alpha' = \alpha$ and set the equilibrium coordinates of initial (L) and final (R) states $\theta_{10}$ and $\theta_{20}$, respectively. The interaction $H_E$ can be expanded as

$$H_E(\theta) = H_{E1} + \frac{\partial H_E}{\partial \theta_{10}}(\theta - \theta_{10}) + \cdots$$

$$= H_{E2} + \frac{\partial H_E}{\partial \theta_{20}}(\theta - \theta_{20}) + \cdots \quad (38)$$

The transition from left conformation to right is divided into three steps: (1) excitation in left valley through

$$\int \psi_{1m\alpha}^+(\theta) H_E(\theta) \psi_{1n\alpha}(\theta) d\theta = \frac{\partial H_E}{\partial \theta_{10}} \int \psi_{1m\alpha}^+(\theta - \theta_{10}) \psi_{1n\alpha} d\theta + \cdots \quad (39)$$

(2) from left to right at the top of barrier through the overlap integral $\int \psi_{2m'\alpha}^+ \psi_{1m\alpha} d\theta$; and (3) transition in the right valley through

$$\int \psi_{2n'\alpha}^+(\theta) H_E(\theta) \psi_{2m'\alpha}(\theta) d\theta = \frac{\partial H_E}{\partial \theta_{20}} \int \psi_{2n'\alpha}^+(\theta - \theta_{20}) \psi_{2m'\alpha} d\theta + \cdots \quad (40)$$

The transition rate in the lowest perturbation

$$W = \frac{2\pi}{\hbar} |T_{fi}|^2 \rho_E$$

$$T_{fi} = \sum_{mm'} \frac{1}{(E_{1n\alpha} - E_{1m\alpha})(E_{1n\alpha} - E_{2m'\alpha})} \int \psi_{2n'\alpha}^+ H_E \psi_{2m'\alpha} d\theta \int \psi_{2m'\alpha}^+ \psi_{1m\alpha} d\theta \int \psi_{1m\alpha}^+ H_E \psi_{1n\alpha} d\theta$$

(41)

## 4. Protein folding rate calculated based on quantum conformational transition

We will calculate protein folding rate based on quantum conformational transition [16][17]

The Hamiltonian of a polypeptide chain is divided into two parts, the kinetic energy $T$ and the potential energy $V$. The all-atom force field for peptides, proteins and organic molecules with fixed bond lengths and bond angles was given in Empirical Conformational Energy Program Package (ECEPP) [18]. The potential function of a polypeptide chain can be expressed by

$$V = \sum \left( \frac{A_{ij}}{r_{ij}^{12}} + \frac{B_{ij}}{r_{ij}^{10}} + \frac{C_{ij}}{r_{ij}^{6}} + \frac{D_{ij}}{r_{ij}} \right) + U_{tor}(\theta_1, ... \theta_N) \quad (42)$$

For fixed bond lengths only the torsion potential $U_{tor}$ should be considered. $U_{tor}$ is assumed to have several minima with respect to each $\theta_i$ and near each minimum the potential can be expressed by a potential of harmonic oscillator. The kinetic energy corresponding to the torsion vibration is

$$T = \sum -\frac{\hbar^2}{2I_{ij}} \frac{\partial}{\partial \theta_i} \frac{\partial}{\partial \theta_j} \quad (43)$$

The kinetic energy can be put in a diagonal form through a linear transformation. So the Hamiltonian is

$$H_1 = \sum -\frac{\hbar^2}{2I_j} \frac{\partial^2}{\partial \theta_j^2} + U_{tor}(\theta_1, ..., \theta_N) \quad (44)$$

here $I_j$ is the inertial moment of the $j$-th mode. To calculate the conformation-transitional rate we consider a more general problem that the dynamical variables of the system are $(\theta, x)$ where $x$ denotes the coordinates of electrons and $\theta = (\theta_1, ..., \theta_N)$, a group of dihedral angles responsible for the conformation change. The wave function of conformation-electron system, $M(\theta, x)$, satisfies Eq (24). As stated in previous section, in adiabatic approximation $M(\theta, x)$ can be expressed as

$$M_{kn\alpha}(\theta, x) = \psi_{kn\alpha}(\theta) \varphi_\alpha(x, \theta) \quad (25a)$$

here $\alpha$ denotes electronic state and $(k,n)$ the conformational and vibration state respectively. For example, in argument-separable case

$$\psi_{kn\alpha}(\theta) = \psi_{k_1, n_1, \alpha_1}(\theta_1) ...... \psi_{k_N, n_N, \alpha_N}(\theta_N) \quad (45)$$

where $\psi_{k_j, n_j, \alpha_j}(\theta_j)$ can approximately be expressed by a wave function of harmonic oscillator with quantum number $n_j$. Note that the harmonic potential has equilibrium position at $\theta_j = \theta_{kj}^{(0)}$ with the corresponding $k_j$-th minimum of potential $E_{kj}$ ($k_j$=1,2,…). By use of the nonadiabaticity operator method and by consideration of $\alpha' = \alpha$ in the protein folding process and the reality of the wave function $\varphi_\alpha$ the conformation transition is described by

$$\langle k'n'\alpha | H' | kn\alpha \rangle = \int \psi^+_{k'n'\alpha}(\theta) \sum_j \{-\frac{\hbar^2}{2I_j} \int \varphi^+_\alpha \frac{\partial^2 \varphi_\alpha}{\partial \theta_j^2} d^3x\} \psi_{kn\alpha}(\theta) d\theta \quad (32a)$$

After thermal average of the initial states the transition rate will take the form

$$W = \frac{2\pi}{\hbar} \sum_{\{n\}} |\langle k'n'\alpha | H' | kn\alpha \rangle|^2 B(\{n\},T) \delta(\sum_j (n'_j \hbar \omega'_j - n_j \hbar \omega_j - \delta E_j)) \quad (46)$$

here $\omega_j$ and $\omega'_j$ are potential parameters of the j-th mode in conformational state $k_j$ and $k_j'$ respectively and $\delta E_j = E_{kj} - E_{k'j}$. $B(\{n\},T)$ denotes the Boltzmann factor (see later). After summing over final states we have

$$W = \frac{2\pi}{\hbar} \sum_{\{n\}} |\langle k'n'\alpha | H' | kn\alpha \rangle|^2 B(\{n\},T) \rho_E \quad (47)$$

Here $\rho_E$ means state density,

$$\rho_E = \frac{\partial n'_1}{\partial E_f}, \qquad E_f = \sum_j (n'_j \hbar \omega'_j + E_{k'j}) \quad (48)$$

If only one torsion angle participates in the conformational transition we call it single-mode or single torsion transition; if several torsion angles participate simultaneously in one step of conformational transition then we call it multi-mode or multi-torsion transition. Consider single-mode case at first. In this case the subscript *j* will be dropped. The Boltzmann factor is given by

$$B(n,T) = e^{-n\beta\hbar\omega}(1 - e^{-\beta\hbar\omega}) \quad (49)$$

The transitional rate

$$W = \frac{2\pi}{\hbar^2 \omega'} I_E I_V \quad (50)$$

$$I_E = \left| \frac{-\hbar^2}{2I} \int \varphi_\alpha \frac{\partial^2}{\partial \theta^2} \varphi_\alpha d^3x \right|^2_{\theta=\theta_0} \quad (51)$$

$$I_V = \sum_n \left| \int \psi^+_{k'n'\alpha}(\theta) \psi_{kn\alpha}(\theta) d\theta \right|^2 B(n,T) \quad (52)$$

$\theta_0$ means the conformational coordinate taking a value of the largest overlap region of vibration functions. Eq. (51) can be estimated roughly by the square of rotational kinetic energy of the frontier electron during conformational change. The quantum number *n'* in (52) is determined by *n* through energy conservation. The overlap integral in Eq (52) was first calculated by Manneback (1951). For simplicity we shall assume the same frequency for initial and final states ($\omega=\omega'$) in the next calculation. After thermal average it gives [19][20]

$$I_V = (\frac{\overline{n}+1}{\overline{n}})^{p/2} J_p(2Q\sqrt{\overline{n}(\overline{n}+1)}) e^{-Q(2\overline{n}+1)} \tag{53}$$

in which

$$\overline{n} = (e^{\beta\hbar\omega} - 1)^{-1}$$

$$Q = I\omega(\delta\theta)^2 / 2\hbar, \qquad p = \frac{\delta E}{\hbar\omega} \tag{54}$$

$\delta\theta = \theta_k^{(0)} - \theta_{k'}^{(0)}$ is the angular displacement and $\delta E = E_k - E_{k'}$ the energy gap between initial and final states ($E_k$ and $E_{k'}$ mean the minimum of the potential in initial and final state respectively). $J_P$ denotes the modified Bessel function [21] and here $p$ is related to the net change in oscillator quantum number. The modified Bessel function is introduced from the expansion

$$\exp\{\frac{x}{2}(y + \frac{1}{y})\} = \sum_{n=-1}^{\infty} y^n J_n(x) \tag{55}$$

By use of the asymptotic formula for Bessel function [22]

$$e^{-z} J_p(z) = (2\pi z)^{-1/2} \exp(-p^2/2z) \qquad \text{for } z \gg 1 \tag{56}$$

$I_V$ can be further simplified. Considering $k_B T / \hbar\omega \approx O(10^1)$ and $\overline{n} \gg 1$ in the range of typical frequency of torsional vibration, it leads to

$$I_V = (2\pi z)^{-1/2} \exp(-\frac{p^2}{2z}) \exp\frac{\delta E}{2k_B T} \tag{57}$$

from Eq (53) where

$$z = (\delta\theta^2) \frac{k_B T}{\hbar^2} I \tag{58}$$

For multi-torsion transition, consider the case of $N$ modes with same frequency $\omega_1 = \omega_2 = ... = \omega_N = \omega$, $\omega'_1 = \omega'_2 = ... = \omega'_N = \omega'$, and $\omega = \omega'$. By using

$$B(\{n\}, T) = \prod_j B(n_j, T) = \prod_j (e^{-n_j \beta\hbar\omega_j}(1 - e^{-\beta\hbar\omega_j})) \tag{59}$$

one has

$$W = \frac{2\pi}{\hbar^2 \omega} I_E \sum_{\{p_j\}} \prod_j I_{Vj} \tag{60}$$

$$I_E = \left| \sum_j \frac{-\hbar^2}{2I_j} \int \varphi_\alpha \frac{\partial^2}{\partial \theta_j^2} \varphi_\alpha d^3 x \right|^2_{\theta_j = \theta_{j0}} \tag{61}$$

$$I_{Vj} = \exp\{-Q_j(2\bar{n}+1)\}(\frac{\bar{n}+1}{\bar{n}})^{p_j/2} J_{p_j}(2Q_j\sqrt{\bar{n}(\bar{n}+1)}) \qquad (62)$$

with

$$Q_j = I_j\omega(\delta\theta_j)^2/2\hbar \qquad (63)$$

and $p_j$ representing the net change in quantum number for oscillator mode $j$ and satisfying the constraint

$$\sum_j p_j = p \equiv \sum \Delta E/\hbar\omega \qquad \Delta E = \sum \delta E_j \qquad (64)$$

in the summation of Eq (60). Here $\delta\theta_j$ is the angular displacement and $\delta E_j$ the energy gap between the initial and final states for the $j$-th mode. Note that Eq (62) takes the same form as (53). To simplify $\sum_{\{p_j\}} \prod_j I_{Vj}$ in Eq (60) we consider two-mode ($N=2$) case at first,

$$\sum_{p_1+p_2=p} I_{V1}I_{V2} = (\frac{\bar{n}+1}{\bar{n}})^{p/2} \times \sum_{p_1}$$
$$\exp\{-Q_1(2\bar{n}+1)\}J_{p_1}(2Q_1\sqrt{\bar{n}(\bar{n}+1)}) \exp\{-Q_2(2\bar{n}+1)\}J_{p-p_1}(2Q_2\sqrt{\bar{n}(\bar{n}+1)}) \qquad (65)$$

By use of the asymptotic formula for Bessel function, Eq (56), we deduce

$$\sum_{p_1} \exp(-z_1)J_{p_1}(z_1)\exp(-z_2)J_{p-p_1}(z_2)$$
$$= \frac{1}{2\pi\sqrt{z_1 z_2}} \int_{-\infty}^{\infty} dp_1 \exp\{-(\frac{p_1^2}{2z_1} + \frac{(p-p_1)^2}{2z_2})\} \qquad (66)$$
$$= \frac{1}{\sqrt{2\pi}} \frac{1}{\sqrt{z_1+z_2}} \exp\{-\frac{p^2}{2(z_1+z_2)}\}$$

Using $\bar{n} \gg 1$ and

$$Z_j = (\delta\theta_j^2)\frac{k_B T}{\hbar^2}I_j = \frac{2k_B T}{\hbar\omega}Q_j \cong 2\bar{n}Q_j \qquad (58a)$$

and inserting (66) into (65) we have

$$\sum_{p_1+p_2=p} I_{V1}I_{V2} = \frac{1}{\sqrt{2\pi}} \frac{1}{\sqrt{z_1+z_2}} \exp\{-\frac{p^2}{2(z_1+z_2)}\} \exp\{\frac{\Delta E}{2k_B T}\} \qquad (67)$$

Eq (67) takes the same form as the single-mode formulas, Eq (57).

Next, by use of the similar deduction, we obtain that if

$$\sum_{p_1+\ldots+p_{n-1}=q} \exp(-z_1)J_{p_1}(z_1)\exp(-z_2)J_{p_2}(z_2)\ldots\exp(-z_{n-1})J_{p_{n-1}}(z_{n-1})$$

(68)

$$=\frac{1}{\sqrt{2\pi}}\frac{1}{\sqrt{z_1+\ldots+z_{n-1}}}\exp\{-\frac{q^2}{2(z_1+\ldots+z_{n-1})}\}$$

holds then it leads to

$$\sum_{p_1+\ldots+p_n=p} \exp(-z_1)J_{p_1}(z_1)\exp(-z_2)J_{p_2}(z_2)\ldots\exp(-z_n)J_{p_n}(z_n)$$

(69)

$$=\frac{1}{\sqrt{2\pi}}\frac{1}{\sqrt{z_1+\ldots+z_n}}\exp\{-\frac{p^2}{2(z_1+\ldots+z_n)}\}$$

So, for multi-mode transition we obtain [16]

$$\sum_{\{p_j\}}\prod_j I_{Vj} = \frac{1}{\sqrt{2\pi}}\exp(\frac{\Delta E}{2k_B T})(\sum Z_j)^{-\frac{1}{2}}\exp(-\frac{p^2}{2\sum Z_j}) \quad (70)$$

Eqs (60)(61) and (70) are the final results of multi-mode transition.

## 5. Results and discussions on protein folding rate

We shall make magnitude estimate on protein folding rate. Eq (51) or (61) can be estimated as follows,

$$I_E = \left\{\sum \frac{\hbar^2}{2I_j}\langle l_j^2\rangle\right\}^2$$

(71)

$$=\frac{\hbar^4}{4}(\sum_j \frac{a_j}{I_j})^2 \qquad a_j = \langle l_j^2\rangle > 0$$

$l_j$ is the $j$-th magnetic quantum number (with respect to $\theta_j$) of electronic wave function $\varphi_\alpha(x,\theta)$, $a_j = <l_j^2> \approx O(1)$, here O(1) means a number in the order of magnitude of 1.

Set $\omega = 6\times 10^{12}\sec^{-1}$, the typical frequency of torsion vibration, $k_B T/\hbar\omega = 6.8$. Taking $|\delta\theta_j|=\pi/2$, $|\delta E_j|/\hbar\omega \approx 10$, $\Delta E/\hbar\omega \approx 10$ one has

$$Z_j = 9.6\times 10^{40} I_j [g\cdot cm^2].$$

By use of the contact interatomic distances in polypeptides, for example, 0.32 nm for C-C, 0.28 nm for C-O, 0.29 nm for C-N, etc one estimates the lower bound of inertia moment $I_j$ near

$10^{-38} \text{g} \cdot \text{cm}^2$. So we can approximate

$$\exp(-\frac{p^2}{2\sum z_j}) = \{\exp(-\frac{p^2}{2\bar{z}})\}^{1/N} \sim 1$$

($\bar{z}$ is the average of $z_j$). Finally we obtain

$$W = 0.37 \times 10^{-87} \exp(\frac{\delta E}{2k_B T}) a^2 I^{-2.5} \sec^{-1} \quad (72)$$

($a = \langle l^2 \rangle \simeq O(1)$, and $I$ in unit of [g cm$^2$]) from (50)(51)(57) for single torsion transition and

$$W = 0.37 \times 10^{-87} \exp\frac{\Delta E}{2k_B T} (\sum I_j)^{-1/2} (\sum \frac{a_j}{I_j})^2 \sec^{-1} \quad (73)$$

($a_j \simeq O(1)$, $I_j$ in unit of [g cm$^2$]) from (60)(61)(70) for multi-mode case. For $N$ mode case with the same parameter in different modes we deduce

$$W = 0.37 \times 10^{-87} \exp(\frac{\Delta E}{2k_B T}) a^2 I^{-5/2} N^{3/2} \sec^{-1} \quad (74)$$

Taking $I = 10^{-37} [g \cdot cm^2]$ to $10^{-36} [g \cdot cm^2]$ we have $W \sim 2 \times 10^5 \sec^{-1}$ to $0.7 \times 10^3 \sec^{-1}$ for single torsion transition. For multi-mode transition the factor $N^{3/2}$ makes the transition rate even faster.

The transition rate depends on energy gap $\sum \delta E_j$ as seen from Eqs. (72) to (74). Consider a transition process from conformation A to B. The transition rate is supposed to be W$_{AB}$. For the inverse process the rate is W$_{BA}$. From the above deduction of the rate we see that two processes are related by $\delta E_j \to -\delta E_j, \delta\theta_j \to -\delta\theta_j, p_j \to -p_j$. So

$$W_{AB}/W_{BA} = \exp(\sum \delta E_j / k_B T) \quad (75)$$

on account of $I_{-p}(z)=I_p(z)$ [22]. If the energy gap in a particular step is much smaller than $k_B T$, then the direction of transition between two conformations in the step is actually determined by the ratio of populations of initial and final states. On the contrary, if $|\sum \delta E_j| \gg k_B T$ then only the process with positive $\sum \delta E_j$ takes place and the inverse one can be neglected.

Eqs. (72) to (74) are the final results usable for numerical estimate of protein folding rate. From Eq (73) we obtain the nucleation time and collapse time in protein folding immediately. For N=4 we estimate the time is microsecond to 0.1 millisecond. The dependence of transitional rate on $N^{3/2}$ given by Eq (74) means the cooperative nature of the transition. The formulas can be used to explain the high rate of the transition of polymers between helix and coil states. The calculated high speed of cooperative transition in polypeptide predicts that the regular secondary structures of α helix and β sheet should be observed in the early stage of folding process.

The dependence of folding rate on inertial moment may be used in the discussion of contact order. The contact order is defined through nonlocal contact number between residues and their sequence distance [5]. The larger the sequence distance of a pair of residues, the more the pair will contribute to the contact order; and as a result, the folding event involving the contact needs a longer transition time. On the other hand, from the deduced dependence of folding rate on inertia moment, $W \sim (\sum I_j)^{-1/2} (\sum \frac{1}{I_j})^2$, we find the rate-limiting factor is $(I_{j,\max})^{-1/2}$.

Therefore, those pairs of residue (or atomic group) with high inertia moment may determine the folding rate basically. Since for two residues that are remote in sequence, the formation of a contact between them may require some atomic group with high inertia moment participating in the transition, we expect that there should exist an inverse relation between contact order and folding rate as indicated by experiments.

By using the dependence of transitional rate on $I^{-2.5}$ one may design a simple model to remove the incorrect folded state. As is well known, how to eliminate wrong folding and guide the peptide chain to its final correctly folded and native functional state is an important problem [23]. Suppose that U is an unfolded state (the population denoted by $N_1$), F is the corresponding folded state (the population denoted by $N_2$) and D - a dead end (incorrect folded state, its population denoted by $N_3$). The transition rates between U and F and between U and D are described by $W$ and $W'$ respectively. Suppose the rates in positive and reverse steps are same. Correspondingly, the master equations are

$$dN_1/dt = W(N_2 - N_1) + W'(N_3 - N_1) + a$$
$$dN_2/dt = W(N_1 - N_2) - b$$
$$dN_3/dt = W'(N_1 - N_3) \qquad (76)$$

$a$ means the transition rate from some source of peptide chain to U , $b$ means the rate from F to final products which are removed from the reaction space as soon as they are produced. We assume that accompanying the increase of $N_3$ the transition from dead end D to unfolded U would trigger the action of 'molecular chaperone'. The supposed 'chaperone' binds to U, increases its inertial moment and in turn, makes the rate $W'$ decreasing. If the molecular weight of the supposed molecule is larger than that of the atomic group participating in conformational transition about two orders of magnitude, then the binding of the supposed chaperone to a atomic group responsible for the transition between U and D would largely lower the transition rate for about five orders following $I^{-2.5}$ law . Evidently, it would cause a large reduction of the number of incorrect folding. So, after addition of the chaperone a large amount of incorrect folded molecules would return to U and jump to F along the correct pathway. Another possibility is: the emergence of chaperone increases the energy gap $\sum \delta E_j$ so that the transition rate of the process from U to D is much reduced than the reverse, as seen from Eq (75). The increase of the gap $\sum \delta E_j$ is equivalent to introducing a change in the barrier between two conformations. So, the addition of a chaperone could increase the activation energy and block the misassembly pathway.

From above discussions we have found that the conformation dynamics of macromolecules furnishes a reasonable ground to understand the time scale of elementary folding mechanism. In the meantime, we have deduced the dependence of the folding rate on molecular parameters. Apart from the dependence of the rate on number of torsion angles and on inertial moment, its temperature dependence is also of general interesting.

Taking the $Z_j$ proportional to $T$ into account (Eq (58a))，from (72) and (73) we obtain the temperature dependence of $W$

$$W \sim \exp(\frac{\Delta E}{2k_B T})T^{-\frac{1}{2}} \tag{77}$$

The relation has a simple classical-mechanical explanation. The conformational transition can be looked as a set of oscillators jumping across a barrier. The oscillators are in thermal equilibrium, obeying normal distribution. The factor $T^{1/2}$ comes from the standard deviation of the oscillator displacement.

As is well known, the rate constant of a typical chemical reaction depends on temperature as $k \sim T \exp(-U/RT)$. The temperature dependence of the rate of torsion transition is near that of a chemical reaction but with an obvious difference in high temperature region as seen from Eq (77). Consider a conformation transition with $\Delta E <0$. The transition takes place from a conformation with lower energy to that with higher energy. The rate increases with temperature which is in resemblance with a typical chemical reaction. However, as the temperature increases continuously to

$$T = |\Delta E|/k_B \tag{78}$$

the transition rate attains a maximum and after that, the rate will decrease. It is expected that the abnormal temperature–dependence of conformation transition will be demonstrated in the future experiments.

To conclude, the conformational dynamics gives us many deep insights into the folding mechanism and provides a useful tool for analyzing and explaining experimental facts on the rate of protein folding.

Acknowledgement. The work was supported by the National Science Foundation of China, project no. 90103030 and 90403010.